\documentclass[aps,pra,twocolumn,epsfig,groupedaddress,showpacs]{revtex4}
\usepackage{graphicx}
\begin{document}

\title{Lifetime Measurement of the $6s$ Level of Rubidium}
\author{E. Gomez, F. Baumer, A. D. Lange, G. D. Sprouse}
\address{Dept. of Physics and Astronomy, State University
of New York at Stony Brook, Stony Brook, NY 11794-3800, U.S.A.}
\author{L. A. Orozco}
\address{Dept. of Physics, University of Maryland,
College Park, MD 20742-4111, U.S.A.}
\date{\today}

\begin{abstract}
We present a lifetime measurements of the $6s$ level of rubidium.
We use a time-correlated single-photon counting technique on two
different samples of rubidium atoms. A vapor cell with variable
rubidium density and a sample of atoms confined and cooled in a
magneto-optical trap. The $5P_{1/2}$ level serves as the resonant
intermediate step for the two step excitation to the $6s$ level.
We detect the decay of the $6s$ level through the cascade
fluorescence of the $5P_{3/2}$ level at 780 nm. The two samples
have different systematic effects, but we obtain consistent
results that averaged give a lifetime of $45.57 \pm 0.17$ ns.

\end{abstract}


\maketitle

\section{Introduction}

This paper presents our measurements of the lifetime of the $6s$
level in rubidium using time-correlated single-photon counting
techniques in two different atomic environments: An atomic vapor
cell, and a magneto-optical trap (MOT). This work complements and
enhances our program of francium spectroscopy and weak interaction
physics.\cite{orozco02} We have previously measured lifetimes in
francium and rubidium only in a
MOT.\cite{simsarian98,grossman00b,aubin03b,aubin04,gomez04,gomez05a}
 We use the same apparatus except for the source of atoms so we
carefully address the different systematic effects that are unique
to each one of them. The present measurements are necessary to
understand systematic effects on the measurement of the equivalent
level in francium, the $8s$ level.\cite{gomez05a}

Measurements of excited state atomic lifetimes in the low-lying
states of the $s$ manifold enhance our understanding of the wave
functions and the importance of correlation corrections in the
theoretical calculations. Relativistic corrections in rubidium are
smaller than in other heavier alkali such as cesium or francium.
This particular level ($6s$) is of primary importance to the
optical PNC measurements that look for the parity forbidden dipole
transition between the ground state and the first excited state
with the same orbital angular momentum.\cite{wood99,guena03}. The
comparison of measurements with theoretical predictions test the
quality of the computed wave functions. The calculation of the
wave functions have now reached new levels of sophistication
\cite{johnson03,ginges04} based on Many Body Perturbation Theory
(MBPT). Those calculations are particularly important in the
interpretation of precision tests of discrete symmetries in atoms
such as Parity Non-Conservation (PNC) \cite{wood99} and Time
Reversal (TR).\cite{khriplovich97}

The paper is structured as follows: We show the relationship
between lifetimes and atomic structure in section \ref{lifetimes}.
Sec. \ref{method} presents the experimental method for the
lifetime measurement with details on the two sources of atoms.
Sec. \ref{data} has the data analysis for the lifetime
measurement. Sec. \ref{results} summarizes the results and
compares them with previous results and theoretical predictions.
Sec. \ref{conclusions} contains the conclusions in the context of
similar measurements in rubidium and francium.

\section{Lifetimes and Atomic Structure}
\label{lifetimes}

The lifetime of an exited atomic state depends on the initial and
final state wave functions and the interaction that connects them.
Since the electromagnetic interaction in atomic physics is well
understood, radiative lifetimes give information on atomic
structure.

The lifetime $\tau$ of an excited state is determined by its
individual decay rates, $1/\tau_i$, to other states by

\begin{equation}
\frac{1}{\tau} = \sum_i \frac{1}{\tau_i}. \label{lifetime}
\end{equation}
The matrix element associated with a partial decay rate ($i$)
between two states connected with an allowed dipole transition in
free space is:

\begin{equation}
\frac{1}{\tau_{i}}= \frac{4}{3}\frac{\omega^{3}}{c^{2}}\alpha
\frac{| \langle J \|r\| J' \rangle |^{2}}{2J'+1} ,
\label{matrixelement}
\end{equation}
where $\omega$ is the transition energy divided by $\hbar$, $c$ is
the speed of light, $\alpha$ is the fine-structure constant, $J'$
and $J$ are respectively, the initial and final state angular
momenta, $\tau_i$ is the excited state partial lifetime, and $|
\langle J \|r\| J' \rangle |$ is the reduced matrix element.

A calculation of the reduced matrix element requires the precise
knowledge of the electronic wave functions involved. The presence
of the radial operator makes the matrix element more sensitive to
contributions of the wave functions at large distances from the
nucleus.

\section{Experimental method}\label{method}

We use time-correlated single-photon counting \cite{oconnor84} to
obtain the lifetime of the $6s$ level in rubidium in a vapor cell
and in a MOT. This two step transition method is well established
both for vapor cells \cite{hoeling96} and MOTs
\cite{aubin04,gomez04} in alkalis.

\begin{figure}
\leavevmode \centering
\includegraphics[width=3.1in]{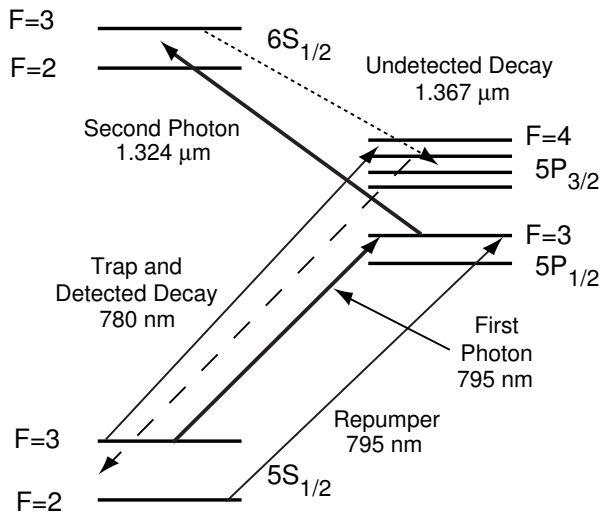}\caption{Energy levels of $^{85}$Rb. The figure
shows the two step excitation (solid thick lines), the
fluorescence detection used in the lifetime measurement (dashed
line) and the undetected fluorescence (dotted line). The trap and
repumper beams (solid thin lines) are only used for the
measurement in the MOT. \label{levels}}
\end{figure}

Figure \ref{levels} shows the energy levels of $^{85}$Rb ($I=5/2$)
relevant to the lifetime measurement. We excite the $6s$ level in
a two-photon resonant process. The first step laser at 795 nm
populates the $5P_{1/2}$ $F = 3$ level and the second step laser
at 1324 nm excites the $5P_{1/2} \rightarrow 6S_{1/2}$ transition.

After we turn off the excitation lasers, the atom returns back to
the ground level using two different decay channels. First, by
emitting a 1324 nm photon it decays back to the $5P_{1/2}$ state
and fluoresces 795 nm light to return to the $5s$ ground level.
This decay path is not used for the $6s$ level lifetime
measurement. The second possible decay channel is the $6s
\rightarrow 5P_{3/2}$ transition followed by the decay to the $5s$
ground level. The 1367 nm fluorescence from the first step of this
decay is unobserved, but we detect 780 nm light from the second
part of the decay. Since the lifetime of the $5P_{3/2}$ state is
well known,\cite{simsarian98,volz96b,gutterres02,heinzen97} it is
possible to extract the $6s$ level lifetime indirectly from the
fluorescence distribution of the second step of the cascading
decay.

We use a high efficiency MOT to capture a sample of rubidium atoms
at a temperature lower than 300 $\mu$K. The MOT is in line with
the Stony Brook Superconducting LINAC and is optimized to trap
francium.\cite{aubin03a} We load the MOT from a rubidium vapor
produced by a dispenser in a glass cell coated with a dry film to
reduce sticking to the walls. The MOT consists of three pairs of
retro-reflected beams, each with 15 mW/cm$^2$ intensity, 3 cm
diameter (1/e intensity) and red detuned 19 MHz from the atomic
resonance. A pair of coils generates a magnetic field gradient of
9 G/cm. The trap contains 10$^4$ atoms with a diameter of 0.5 mm
and a typical lifetime between 5 and 10 s.

We perform the lifetime measurement also in a vapor cell. The cell
is under vacuum (approximately $10^{-5}$ Pa) and a dispenser
provides the rubidium for the measurement. The cell is uncoated
and we use no buffer gas. We keep the rubidium density low to
avoid collisional quenching and radiation trapping effects. The
typical mean free path of the atoms in the cell is more than 30 m.
We can apply a uniform magnetic field to the cell in contrast to
the permanent magnetic field gradient in the MOT.

\subsection{Lasers System}

\begin{figure}
\leavevmode \centering
\includegraphics[width=3.1in]{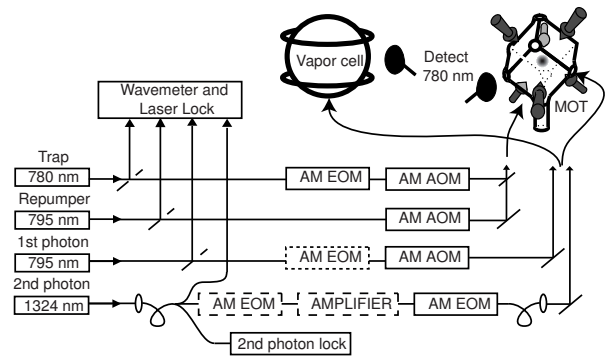}
\caption{Experimental setup. The trap and repumper beams are not
used for the vapor cell measurement. The dotted (dashed) boxes are
not used for the MOT (vapor) experiment. AM EOM=amplitud modulated
electro-optic modulator, AM AOM=amplitude modulated acousto-optic
modulator \label{lasers}}
\end{figure}

Figure \ref{lasers} shows the experimental setup. We use a
Coherent 899-21 titanium-sapphire (Ti:sapph) laser at 780 nm
between the $5s$ $F=3$ and the $5P_{3/2}$ $F=4$ levels to trap and
cool the atoms. A Coherent 899-21 Ti:sapph laser at 795 nm between
the $5s$ $F=2$ and the $5P_{1/2}$ $F=3$ repumps any atom that
falls out of the cycling transition. The trap and repumper beams
are not necessary for the vapor cell measurement. We excite the
atoms to the $6s$ level via a two step transition. A Coherent
899-01 (or a Coherent 899:21 for the vapor cell) at 795 nm between
the $5s$ $F=3$ and the $5P_{1/2}$ $F=3$ levels makes the first
step to the $5P_{1/2}$ level and a EOSI 2010 diode laser at 1.324
$\mu$m completes the transition to the $6s$ level.

\begin{figure}
\leavevmode \centering
\includegraphics[width=3.1in]{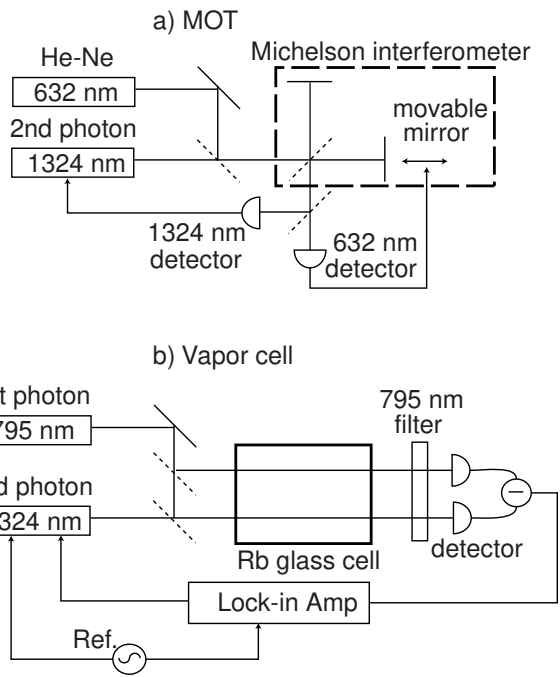}\caption{Block diagram for the 1324 nm
second step laser locking system for the measurement: a) with the
MOT and b) with the vapor cell. \label{lock}}
\end{figure}

A Burleigh WA-1500 wavemeter monitors the wavelength of all the
lasers to about $\pm 0.001$ cm$^{-1}$. We lock the trap, repumper
and first step laser with a transfer lock.\cite{zhao98} We use a
different method to lock the second step laser on the vapor cell
and on the MOT measurements as described below. Figure \ref{lock}
shows both locking methods for the second step laser.

{\bf Second step laser lock for the MOT.} We transfer the
stability of the Helium-Neon laser (Melles-Griot He-Ne 05-STP-901)
used on the lock of the other lasers \cite{zhao98} to lock a
Michelson interferometer with a 200 MHz free spectral range. The
1324 nm light propagates collinearly with the He-Ne light inside
the interferometer. We detect the output of the locked
interferometer at 1324 nm and use that signal to lock the second
step laser to a side of the fringe, which allows us to set it on
the atomic resonance by an offset adjustment. We achieve a drift
smaller than 5 MHz per hour.

{\bf Second step laser lock for the vapor cell.} We use two photon
modulation transfer spectroscopy to lock the second step laser. We
split the first step laser at 795 nm into two beams and send both
of them through a separate rubidium glass cell. The room
temperature vapor cell is uncoated and the beam propagates
approximately 7 cm inside the cell, which leads to a typical
absorption of about 15 \%. We collinearly apply the 1.324 $\mu$m
second step laser to one of the beams to excite the second
transition. If the laser is resonant to the two step transition it
results in less absorption of the 795 nm light. Two beams and
balanced detection allow common mode noise rejection to increase
the signal to noise ratio. We use the internal piezoelectric
crystal of the 1.324 $\mu$m diode laser to dither its frequency at
1.3 kHz and perform lock-in detection of the 795 nm light
transmission through the cell. We feedback the resulting error
signal to the second step laser to lock it with comparable
performance to that of the Michelson lock.

We send and retro-reflect 1.5 mW of the first step laser power to
the trap region. We focus it to an spot size approximately equal
to the trap size. The second step laser travels at an small angle
with respect to the first step laser and we also focus it to match
the trap size with a power of 1 mW. For the vapor cell measurement
the excitation beams propagate collinearly inside of the glass
cell. We send 100 $\mu$W of power at 795 nm and 0.4 mW of power at
1.324 $\mu$m and focus it to a spot size of about 0.6 mm.

\subsection{Timing}

\begin{figure}
\leavevmode \centering
\includegraphics[width=8.6cm]{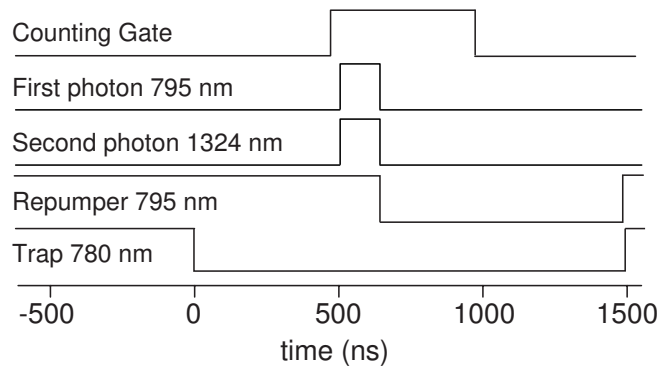}\caption{Timing diagram for the $6s$ level
excitation and decay cycle. The cycle repetition rate is 100 kHz.
There is no trap nor repumper for the vapor cell measurement.
\label{timing}}
\end{figure}

Figure \ref{timing} displays the timing sequence for the
excitation and decay cycle. We apply the two step excitation 500
ns after the trap laser turnoff for a duration of 100 ns. We turn
off the repumper at the same time as the two step lasers to look
for the fluorescence decay of the atoms. The counting electronics
are sensitive for 500 ns to record the excitation and decay
signal. The cycle repetition rate is 100 kHz. The trap and
repumper pulses are not necessary for the vapor cell measurement.

{\bf Modulation for the MOT.} We turn the trap laser on and off
with an electro optic modulator (EOM) (Gs${\rm \ddot{a}}$nger
LM0202) and an acousto-optic modulator (AOM) (Crystal Technology
3200-144) as shown in Fig. \ref{lasers}. The EOM gives a fast
turnoff and the AOM improves the long term on/off ratio. The
combination of the two gives an extinction ratio of better than
1600:1 after 500 ns. We modulate the repumper and first step laser
with an AOM (Crystal Technology 3200-144) with an on/off ratio of
109:1 and 26:1 respectively after 30 ns. We couple the 1324 nm
light into a single mode optical fiber. We use a combination of a
10 Gbits/s lithium niobate electro-optic fiber modulator (Lucent
Technologies 2623N) followed by an fiber amplifier (InPhenix
IPSAD1301) and a second identical fiber modulator to turn the
laser on and off. The combination of the three elements gives an
on/off ratio better than 1000:1 after 20 ns.

{\bf Modulation for the vapor cell.} For the vapor cell
measurement we do not need the trap and repumper beams and we
install the EOM on the first step laser path instead. In this case
we achieve an extinction ratio of better than 100:1 in 30 ns. We
reduce any radio frequency (RF) emission generated by the fast
turnoff of the EOM placing it in a separate room inside a metal
cage. We use only the last fiber modulator for the 1324 nm laser
and we achieve an on/off ratio of 110:1 after 20 ns. Figure
\ref{lasers} shows an schematic diagram of the modulation of all
the lasers.

\subsection{Imaging system}

A 1:1 imaging system (f/3.9) collects the MOT fluorescence photons
onto a charge coupled device (CCD) camera (Roper Scientific,
MicroMax 1300YHS-DIF). We monitor the trap with the use of an
interference filter at 780 nm in front of the camera. A
beam-splitter in the imaging system sends 50 \% of the light onto
a photomultiplier tube (PMT) (Hamamatsu R636). An interference
filter at 780 nm in front of the PMT reduces the background light
other than fluorescence from the cascade through the $5P_{3/2}$
level decay back to the ground state $5S_{1/2}$.

The 1:1 (f/2.4) imaging system for the vapor cell is perpendicular
to the excitation beams and collects the fluorescence from the
decaying atoms to a PMT (Hamamatsu R636) with a 780 nm
interference filter in front of it.

\subsection{Detection electronics}

\begin{figure}
\leavevmode \centering
\includegraphics[width=3.1in]{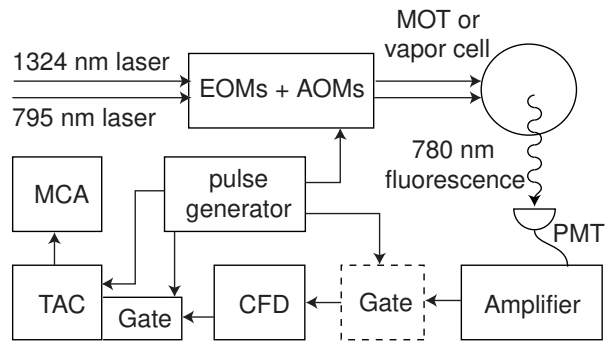} \caption{Block diagram of the processing
electronics. The dashed gate is not used for the vapor cell
experiment. PMT=photomultiplier tube, CFD=constant fraction
discriminator, TAC=time to amplitude converter, MCA=multichannel
analyzer. \label{imaging}}
\end{figure}

Figure \ref{imaging} shows the electronic processing of the
detected fluorescence. When the PMT detects a photon, it generates
a variable amplitude electronic pulse. We first amplify the signal
with an Ortec AN106/N amplifier, we send it to a linear gate
(EG$\&$G LG101/N, not used for the vapor cell measurement) and
then to an Ortec 934 constant fraction discriminator (CFD). The
output of the CFD goes to a gated time-to-amplitude converter
(TAC) (Ortec 467). The gating is done directly on the TAC for the
MOT measurement and with an external gate (Ortec LG101/N) for the
vapor cell measurement. The photon pulse starts the TAC and we
stop it with an electronically generated pulse that has a fixed
time delay with respect to the excitation laser pulse. Starting
the TAC with a fluorescence photon eliminates the accumulation of
counts from cycles with no detected photons. We use a multichannel
analyzer (MCA) (EG$\&$G Trump-8k) to produce a histogram of the
events showing directly the exponential decay. A Berkeley
Nucleonics Corporation BNC 8010 pulse generator is the master
clock with slaved pulse and signal generators to provide the
timing sequence for the measurement.

\section{Data Analysis}\label{data}

We extract the atomic radiative lifetime from the measured
fluorescence decay. We take sets of data for about 600 s, that are
individually processed, added together, and fitted. The $6s$ level
decays through the $5p$ states to the $5s$ level as shown in
Fig.~\ref{levels} and we detect the indirect decay through the
$5P_{3/2}$ state. The fitting function is a sum of two
exponentials, a background and an sloping background coming from
the trap laser turnoff
\begin{eqnarray}
S_{6s}&=&  A_{6s} \exp\left(-\frac{t}{\tau_{6s}}\right) + A_{5p}
\exp\left(-\frac{t}{\tau_{5p}}\right)  \\ \nonumber & & + A_B
+A_St \label{ffct},
\end{eqnarray}
where $\tau_{5p}$ is the known lifetime of the $5P_{3/2}$ state
and $\tau_{6s}$ the lifetime we want to extract, $A_{6s}$ and
$A_{5p}$ are the coefficients of the exponentials while $A_B$ and
$A_S$ characterize the background. $A_S=0$ for the vapor cell
measurement since there is no trap laser. Equation \ref{ffct}
reduces to a single exponential if one waits long enough after the
turnoff to take data.\cite{diberardino98} We instead maximize the
statistics by including both exponentials in our analysis.

\begin{figure}
\leavevmode \centering
\includegraphics[width=3.1in]{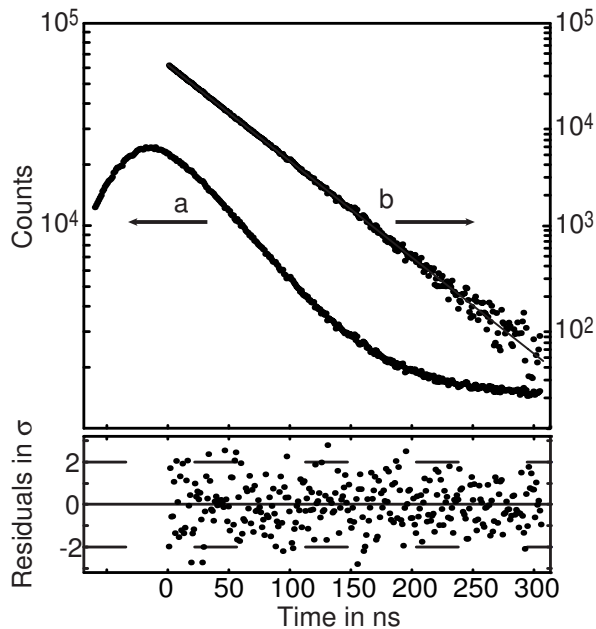} \caption{Decay curve of the $6s$ level through the
$5P_{3/2}$ state in the vapor cell with fit and residuals. The
upper plot shows a) the raw arrival time histogram data for the
vapor cell measurement and b) the data after the pile-up
correction and substraction of background and the $5P_{3/2}$
decay. The lower plot shows the normalized residuals.
$\chi_{\nu}^2$=0.98 \label{decay}}
\end{figure}

The first part in the data analysis requires a pile-up correction,
that accounts for the preferential counting of early
events.\cite{oconnor84} If $N_i$ is the number of counts in the
MCA channel $i$, and $N_E$ is the total number of excitation
cycles, then $N_i'$ is the corrected number of counts in channel
$i$ given by

\begin{equation}
N_i'=\frac{N_i}{1-\frac{1}{N_E}\sum_{j=1}^{i-1}N_j} \label{pileup}
\end{equation}
As low count rates keep this correction small, we collect data
with a small number of fluorescence photons. We typically count
one photon every 100 cycles, which corresponds to a correction in
the fitted lifetime by 0.1 \%.

We perform a nonlinear least square fit using Microcal$^{\rm TM}$
Origin$^{\rm TM}$ Version 5.0 to find the fitting parameters that
produce the smallest $\chi^2$. We have tested the software
independently to validate and understand its results. Figure
\ref{decay} shows an example of the fitting procedure. We start
the fit 20 ns (25 ns for the vapor cell measurement) after both
excitation lasers are off. The fitting function describes the data
well, and the reduced $\chi_{\nu}^2$ of this particular decay is
0.98. The average $\chi_{\nu}^2$ for the MOT data is 1.05 $\pm$
0.06 and for the vapor cell 1.09 $\pm$ 0.08. A discrete Fourier
transform of the residuals shows no structure.

The MOT and vapor cell measurements have some characteristics that
are intrinsically different between them. The MOT cools the atoms
to a very low temperature, while the atoms in the vapor cell have
temperatures corresponding to Doppler shifts significantly larger
than the natural linewidth of the transitions of interest. The MOT
has magnetic field gradient always present which can be removed
for the vapor cell. These differences generate systematic effects
that are intrinsic to each system. We classify the systematic
effects into two groups, common effects to both the MOT and the
vapor cell measurement and distinct effects that appear in a
different way in both systems.

\subsection{Common systematic effects}

\subsubsection{Truncation uncertainty}

The truncation uncertainty takes into account the variation of the
fitted lifetime depending on the start and end points of the fit.
We find no statistically significant change in the obtained
lifetime when we change both the initial or final point of the
fit.

\subsubsection{Time calibration}

We calibrate the scale on the MCA by sending a series of
artificial start and stop pulses with known delay directly to the
TAC either from one of the DG535 pulse generators or from two
DG535 pulse generators triggered by a frequency synthesizer
(Rockland 5600). The linear fit to the resulting data contributes
a $\pm$ 0.01 \% to the lifetime uncertainty and it is consistent
using either source of calibration pulses.

\subsubsection{TAC and MCA response nonuniformity}

A nonuniformity in the TAC or MCA results in different bins of the
MCA having different response sensitivities that alters the
obtained lifetime. We measure the height uniformity of the TAC/MCA
system by triggering the PMT with random photons from a white
light source. The nonuniformity of the TAC/MCA contribute a 0.11
\% (0.02 \% for the vapor cell measurement) to the uncertainty of
the lifetime. The need for additional electronic gates in the MOT
measurement increases the nonuniformity in the electronics
response. The response changes slightly between days. It shows up
in the data as an additional noise contribution that is
responsible for the deviation of the $\chi_{\nu}^2$ from one. The
effect is already included in this uncertainty contribution.

\subsubsection{Imperfect laser turnoff}

An imperfect turnoff of the excitation lasers influences the
observed decay signal. If the second step laser is not completely
turned off during the decay, atoms that decay to the $5P_{1/2}$
state can be reexcited to the $6s$ level. We test the turnoff by
comparing the lifetime obtained when we leave the first step laser
continuously on or off during the decay. We establish a limit of
0.07 \% for the effect of the imperfect turnoff of the second step
laser in the MOT measurement.

We use two modulators for the second step laser on the MOT
measurement to achieve a large on/off ratio. The vapor cell
measurement uses only one modulator and we apply a correction to
the obtained lifetime.

The appendix presents the model used to calculate the effect of an
imperfect turnoff on the lifetime. We solve the rate equations and
compare the result obtained with a perfect turnoff to the
experimental one. We find a correction to the lifetime of - 0.11
ns and a sensitivity analysis of the parameters of the model gives
an uncertainty contribution of $\pm$ 0.1 \%.

\subsubsection{Initial state conditions}

The lifetime should be independent of the populations of the
different levels at the beginning of the decay. We can change the
initial state conditions by changing the power of the first or
second step laser pulses, or their duration.  We calculate the
linear correlation coefficient and its integral probability to
study the correlation between the obtained lifetime and the
external variable. The larger the integral probability is, the
less likely it is that a correlation exists \cite{bevington}.

We vary the power of the first and second step lasers by more than
an order of magnitude. In the vapor cell measurement we also
changed the excitation pulse duration between 50 ns and 200 ns. In
all the above cases we found results consistent with no
correlation.

\subsection{Distinct systematic effects}

\subsubsection{Radiation trapping}

The fluorescence photons from atoms decaying from the $5P_{3/2}$
state to the $5s$ ground level can excite other atoms that are in
the ground state and take them to the $5P_{3/2}$ level, the so
called radiation trapping.\cite{holstein47,beeler03} Traps are
very convenient systems to make measurements because the atoms are
collected into an small region of space. The effect of radiation
trapping is small in a low density trap because the distance $d$
corresponds to the diameter of the trap which is rather small.
Vapor cells on the other hand may have a large density, but
Doppler shifts reduce the effective number of atoms that could
reabsorb the fluorescence.

The probability of a photon being absorbed by an atom can be
deduced from Beer's law
\begin{equation}
I(d) = I_0 \exp(-\alpha d),
\end{equation}
where $I_0$ is the initial intensity and $I(d)$ is the intensity
after traversing the distance $d$ and $\alpha=\sigma n$ represents
the absorption coefficient, with $n$ the density of the atomic
vapor and $\sigma=3 \lambda^2 f/2\pi$ the absorption cross
section. $f=0.69$ is the oscillator strength of the corresponding
transition. For small absorption the probability $P$ of a photon
being absorbed by the atoms is $P = \sigma n\, d$.

The rate equation for the $5P_{3/2}$ level in the presence of
radiation trapping after the lasers turnoff is given to first
order by
\begin{equation}
\frac{d N_{5p}}{dt} = B_{5p}\frac{N_{6s}}{\tau_{6s}}-
\frac{N_{5p}}{\tau_{5p}} + \frac{N_{5p}}{\tau_{5p}} P,
\end{equation}
with $N_{5p}$ and $N_{6s}$ the number of atoms in the $5P_{3/2}$
and $6s$ levels and $B_{5p}$ the branching ratio from the $6s$
level to the $5P_{3/2}$. In this approximation the change in the
lifetime of the $5P_{3/2}$ level is directly proportional to $P$.
As our fitting process assumes a fixed $5P_{3/2}$ state lifetime,
the change due to radiation trapping propagates linearly to the
measured $6s$ level lifetime.

We vary the density of rubidium atoms to look for variations in
the lifetime. We change the density by changing the number of
atoms in the MOT or by compressing the trap with an increase the
magnetic field gradient. We establish a limit of 0.01 \% for the
effect of radiation trapping on the lifetime.

A rubidium dispenser (SAES) provides the atoms in the vapor cell
measurement. A zirconium alloy matrix releases rubidium atoms when
it is heated by a DC current. The desorption follows an Arrhenius
process that requires a certain activation energy $E_a$. The
atomic density in the cell follows the relation
\begin{equation}
n(T) \propto \exp\left(\frac{k_B T}{E_a}\right),
\end{equation}
where $k_B$ is the Boltzmann constant and $T$ the temperature in
K. Measurements of the pressure on the cell confirm this
relationship.

\begin{figure}
\leavevmode \centering
\includegraphics[width=3.1in]{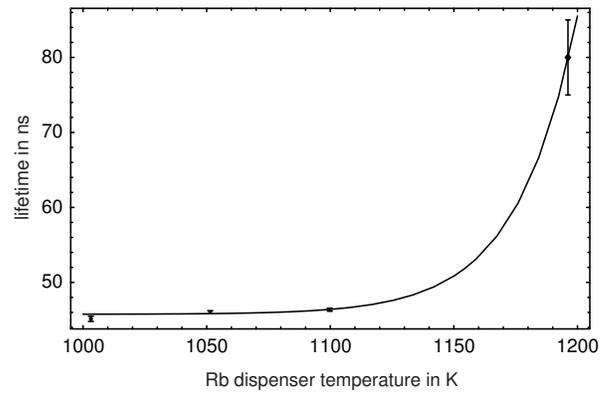} \caption{Temperature dependence of the measured
$6s$ level lifetime together with a fit. \label{arrp}}
\end{figure}

We compare the lifetime obtained at different dispenser
temperatures and extrapolate the observed dependency to zero
temperature to obtain the correction to the lifetime. The fitting
function for the temperature dependence of the observed lifetime
is
\begin{equation}\label{tfit}
\tau(T) = \tau_0 + A \exp(T/B),
\end{equation}
where $A$, $B$ and the corrected lifetime $\tau_0$ are fitting
constants. Figure \ref{arrp} shows the measured data and the
obtained fit. The correction of the $6s$ level lifetime due to
radiation trapping amounts $-0.28$ ns and generates an uncertainty
contribution of 0.38 \% becoming the dominant uncertainty
contribution in the vapor cell measurement.

\subsubsection{Magnetic field}

The presence of a magnetic field can influence the measured
lifetime through quantum beats between Zeeman sublevels. In ideal
conditions the atoms in a MOT sit at the zero of the magnetic
field gradient but in reality an imbalance of the power of the
retro-reflected beams may displace the trap from this ideal
position. The use of a cascade decay reduces the appearance of
quantum beats.\cite{hoeling96,knight80}

We quantify the magnetic sensitivity of the lifetime measurement
in two ways. First we perform the measurement at different
magnetic field gradients. We do not observe a correlation with the
magnetic field gradient and we establish a limit on the effect of
the magnetic field on the lifetime of 0.17 \%. This limit is
consistent with measurements done in the vapor cell. Alternatively
we keep the magnetic field environment fixed and we displace the
trap by creating an imbalance on the laser beams that interact
with it. We can imbalance the trap beams by inserting a piece of
glass in front of one of the retro-reflection mirrors, and
repeating the same for all three axis. Changing the power of the
first step laser beam also displaces the trap. The change in the
obtained lifetime in all cases is consistent with statistical
fluctuations.

The measurement in the vapor cell does not require a magnetic
field gradient. We generate an homogeneous magnetic field with two
coils arranged close to a Helmholtz configuration with its axes in
the same direction as the excitation laser beams. The magnetic
field homogeneity in the excitation region is $\pm$ 0.35 \%.

\begin{figure}
\leavevmode \centering
\includegraphics[width=3.1in]{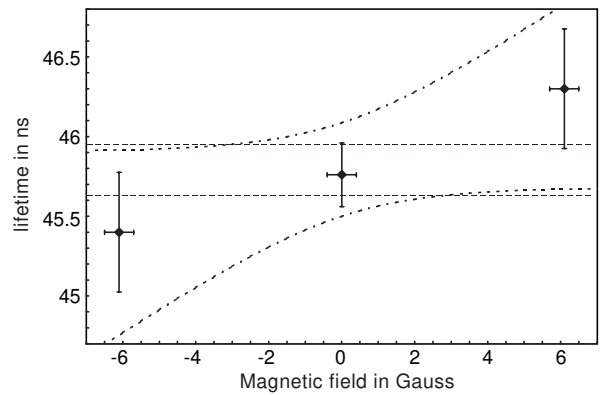} \caption{Magnetic field dependence of
measured lifetime in the vapor cell. The plot contains the
1$\sigma$ region for the linear fit to the data (dotted line) and
the error on the average assuming no linear dependence (dashed
line). \label{magnetic}}
\end{figure}

Figure \ref{magnetic} shows the measured $6s$ level lifetime in
the vapor cell as a function of the external uniform magnetic
field. We can control the magnetic field to $\pm 0.2$ Gauss due to
the earth magnetic field and other magnetic sources. The
correlation coefficient for the three data points is consistent
with no correlation. The dotted line shows the 68\% confidence
band (1$\sigma$) on the linear fit for a $t$ distribution with one
degree of freedom. The dashed line gives the error on the average
assuming no linear dependence. The figure shows that both regions
are consistent with each other. We limit the contribution to the
uncertainty due to magnetic fields to $\pm$ $0.07$ \% using the
linear fit to the data.

\subsubsection{Collisional quenching}

Inelastic collisions can modify the lifetime of an excited state.
The velocity of atoms in the MOT (vapor cell) is smaller than 0.1
m/s (250 m/s). The mean free path of the atoms in the MOT (vapor
cell) is about 10$^4$ m (10 m) the effect of collisions is
negligible for both samples.
\subsubsection{Time of flight}

A moving atom can escape the imaging area before it decays. The
effect is more evident in the vapor cell where the atoms have
larger speeds. The MOT (vapor cell) imaging area has a 1 mm (5.5
mm) diameter aperture. It takes an atom on average 10$^5$ (200)
lifetimes to leave the MOT (vapor cell) imaging area. A
Monte-Carlo simulation shows no statistically significant change
in the lifetime for the vapor cell measurement.

\subsubsection{Background slope}

The detected light at 780 nm comes from the cascade decay from the
$6s$ level, but also from scattered light from the trap laser. We
turn off the trap laser 500 ns before the two step excitation, but
the remanent of the trap laser light appear in the signal as an
slope in the background. Fits to files with and without decaying
atoms give consistent values for the background slope. We compare
the lifetime obtained when we leave the background slope as a free
parameter or when we fix it to the value obtained without two step
excitation and obtain an uncertainty contribution of $\pm$ 0.22
\%. This contribution does not appear in the vapor cell
measurement since there is no trap laser.

\subsubsection{PMT response}

We look for saturation effects in the PMT due to the trap light by
comparing the PMT response to that of a fast photodiode which is
not subject to saturation. We replaced the two step excitation by
a light pulse at 780 nm. Comparison of the detected pulse in the
PMT and in the fast photodiode gives a maximum contribution to the
uncertainty of $\pm$ 0.21 \%. This effect is not present in the
vapor cell measurement.

\section{Results and comparison with theory}
\label{results}

The lifetime of the $6s$ level depends on the value of the
lifetime of the $5P_{3/2}$ state. The uncertainty of the
$5P_{3/2}$ state lifetime propagates to the uncertainty of the
$6s$ level lifetime. Using Bayesian statistics the contribution to
the uncertainty ($\sigma_{B_{6s}}$) due to the $5P_{3/2}$ level is
given by\cite{aubin04}

\begin{equation}
\sigma_{B_{6s}} = \frac{d\tau_{6s}(\tau')}{d\tau'} \;\sigma_{5p}.
\end{equation}
We assume different values $\tau'$ for the lifetime of
$\tau_{5P_{3/2}}$ and include them in the fitting function (Eq.
\ref{ffct}) to obtain the dependence of $\tau_{6s}(\tau')$.

The $5P_{3/2}$ level lifetime has been measured previously by
several groups (see Table \ref{5presults}). We take the weighted
average of all the measurements. We add the systematical
contributions of the respective errors in quadrature and calculate
the total statistical contribution according to $\sigma_{stat}^2 =
(\sum 1/ \sigma_i^2)^{-1} $ where $\sigma_i$ is the statistical
uncertainty of a single result. The $5P_{3/2}$ state lifetime of
26.23(9) ns gives a Bayesian error for the $6s$ level lifetime of
0.17 \% (0.11 \% for the vapor cell).

\begin{table}[h]
\leavevmode \centering
\renewcommand{\arraystretch}{1.1}
\begin{tabular}{lr}
 Author      & ~~~~lifetime [ns] \\ \hline
Simsarian {\it et al.} \cite{simsarian98} &26.20(9) \\
Volz {\it et al.} \cite{volz96b} & 26.24(4) \\
Gutterres {\it et al.} \cite{gutterres02} &26.25(8) \\
Heinzen {\it et al.} \cite{heinzen97} &26.23(6) \\
\end{tabular}
\caption{Experimental results for the rubidium $5P_{3/2}$
lifetime}\label{5presults}
\end{table}

Tables \ref{errorbudgetvc} and \ref{errorbudgetmot} contain the
error budget and the corrections for the $6s$ level lifetime
measurement in the vapor cell and in the MOT. The error is
dominated by the statistical uncertainty in the MOT and by
radiation trapping in the vapor cell. The vapor cell corrections
change the result by -0.9 \%. We obtain a lifetime of 45.64 $\pm$
0.22 ns in the vapor cell and 45.48 $\pm$ 0.25 ns in the MOT for
the $6s$ level lifetime. Both results are consistent with each
other and the average gives 45.57 $\pm$ 0.17 ns.

\begin{table}
\renewcommand{\arraystretch}{1.3}\centering
\begin{tabular}{lrr}
& Correction& ~~~~\% error \\ \hline

Time calibration&&$\pm$0.01 \\
Bayesian error&&$\pm$0.11 \\
TAC/MCA nonuniformity~~~~&&$\pm$0.02 \\
Imperfect laser turnoff&-0.11 ns&$\pm$0.10 \\
Radiation trapping&-0.28 ns&$\pm$0.38 \\
Magnetic Field&&$\pm$0.07 \\
Statistical error &&$\pm$0.24 \\ \hline
{\bf Total} & &{\bf$\pm$0.48} \\
\end{tabular}
\vspace{0.5cm}\caption{Error budget and corrections for the $6s$
level lifetime measurement in the vapor cell}
\label{errorbudgetvc}
\end{table}

\begin{table}
\renewcommand{\arraystretch}{1.3}\centering
\begin{tabular}{lr}
& \% error \\ \hline

Time calibration&$\pm$0.01 \\
TAC/MCA nonuniformity~~~~&$\pm$0.11 \\
Radiation trapping&$\pm$0.01 \\
Imperfect laser turnoff&$\pm$0.07 \\
Magnetic Field&$\pm$0.17 \\
Background slope&$\pm$0.22 \\
PMT response&$\pm$0.21 \\
Bayesian error&$\pm$0.17 \\
Statistical error &$\pm$0.38 \\ \hline
{\bf Total} &{\bf$\pm$0.56} \\
\end{tabular}
\vspace{0.5cm}\caption{Error budget for the $6s$ level lifetime
measurement in the MOT} \label{errorbudgetmot}
\end{table}

\begin{figure}
\leavevmode \centering
\includegraphics[width=3.1in]{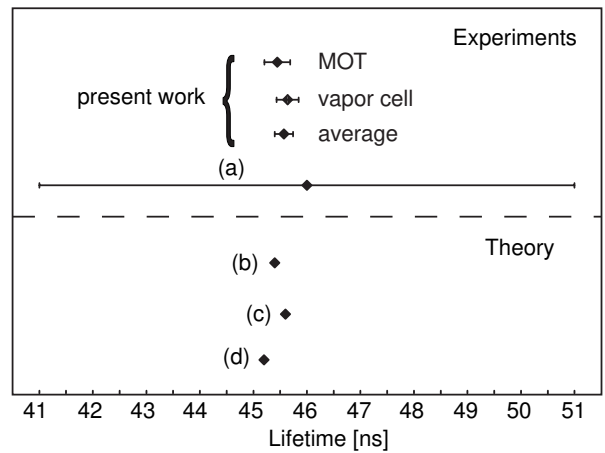}
\caption{Comparison of the present measurement of the $6s$ level
lifetime with previous experimental results (a) by J. Marek
\textit{et al.},\protect{\cite{marek80}} and theoretical
predictions by (b) M. S. Safronova \textit{et
al.},\protect{\cite{safronova04}} (c) W. R. Johnson \textit{et
al.}\protect{\cite{johnson96}} and (d) C. E. Theodosiou \textit{et
al.}.\protect{\cite{theodosiou84}}} \label{comparison}
\end{figure}

Figure \ref{comparison} compares the obtained $6s$ level lifetime
with: (a) a previous measurement by Marek \textit{et
al.},\cite{marek80} (b) and (c) are \textit{ab initio} many body
perturbation theory calculation by Safronova \textit{et
al.}\cite{safronova04} and W. R. Johnson \textit{et
al.}\cite{johnson96} and (d) is  a semiempirical prediction using
a one electron effective potential model by C. E. Theodosiou
\textit{et al.}\cite{theodosiou84}.

\section{Conclusions}
\label{conclusions}

The consistency of lifetime results using two different sources of
atoms demonstrates the understanding achieved in the systematic
effects present in this technique. The MOT source advantages as
far as density, temperature, and radiation trapping compared to a
room temperature cell; however the intrinsic magnetic gradient has
been a source of concern. This work sets a lower limit on their
influence by doing the measurement with the two atomic sources.
Statistics limit the MOT results, while the radiation trapping is
the limiting factor in the glass cell. This measurement improves
previous determinations by a factor of thirty.

Our measurement of the lifetime of the first excited state in the
$s$ manifold of rubidium has reached a precision of 0.3 \% which
permits a very careful tests of the calculated matrix elements
that contribute to the lifetime of the $6s$ state. These {\it ab
initio} MBPT calculations represent the state of the art in atomic
structure calculations and require a very thorough understanding
of the electron correlations and many other subtle effects that
add and subtract in the expansions. The same techniques applied
for rubidium are used in cesium and the agreement between
experiment and theory enhances the confidence in the established
calculation techniques to provide the necessary information such
as matrix elements for the extraction of weak interaction
parameters from the PNC measurements in cesium and in future
measurement in other alkali atoms.

\begin{acknowledgments}
Work supported by NSF. E.G. acknowledges support from CONACYT and
the authors thank the personnel at the Nuclear Structure
Laboratory at Stony Brook for their support and J. Gripp, and J.
E. Simsarian and Bill Minford for equipment loans for this work.
\end{acknowledgments}

\section{Appendix}

We calculate the effect of the imperfect laser turnoff by
including it in the rate equations as a small perturbation and
compare the result to the unperturbed solution. For simplicity we
rename the $6s$, $5P_{3/2}$, $5P_{1/2}$ and $5s$ levels as $1$,
$2$, $3$, and $4$ respectively. The rate equations with the two
step excitation on are
\begin{eqnarray}
\dot{N_1} &=& - \frac{N_1}{\tau_1} + e_1 N_3 \nonumber \\
\dot{N_2} &=& - \frac{N_2}{\tau_2} + B_2 \frac{N_1}{\tau_1} \nonumber \\
\dot{N_3} &=& - \frac{N_3}{\tau_3} + e_3 N_4 + B_3
\frac{N_1}{\tau_1} - e_1 N_3\label{e3},
\end{eqnarray}
where $N_i$ represents the number of atoms in state $i$ with
$N_1+N_2+N_3+N_4=N$ the total number of atoms, $\tau_i$ is the
lifetime of state $i$, $e_i$ stands for the excitation to state
$i$, and $B_i$ is the branching ratio from state 1 to state $i$.

We take the steady state solution of Eq. \ref{e3} as the initial
state for the decay. The solution of Eq. \ref{e3} with $e_1=e_3=0$
corresponds to the case of a perfect on/off ratio.

To calculate the perturbation introduced by the imperfect turnoff
we use the unperturbed solution for $N_3(t)$ on the other two
equations in the following way
\begin{eqnarray}
\dot{N_1} &=& - \frac{N_1}{\tau_1} + f(t)\, e1\, N_3 \nonumber \\
\dot{N_2} &=& - \frac{N_2}{\tau_2} + B_2
\frac{N_1}{\tau_1}\label{d5},
\end{eqnarray}
with $f(t)$ the second step laser turnoff which is well
represented by a Gaussian function ($\Delta t \approx5$ ns) plus a
constant level
\begin{equation}
f(t) = (1 - a) \exp\left(\frac{t}{\Delta t}\right)^2 +
a\label{turnofffunction},
\end{equation}
where $a$ is the on/off ratio.

We use the analytic solution of Eq. \ref{d5} for $N_2(t)$ to
generate a decay sample that we process in the same way as real
data to obtain a lifetime for the $6s$ level.

We include the following parameters in the model: For the
lifetimes of the $5p$ levels we use the results
\cite{simsarian98,volz96b,gutterres02,heinzen97} (as in Table
\ref{5presults}) and for the branching ratios $B_i$ we use the
theoretical values calculated by Safronova \textit{et
al.}\cite{safronova04} We take $e_3=1/\tau_3$ and to obtain $e_1$
we compare the observed number of counts when the excitation
reaches steady state to $N_2(0)$ in the model.

The most likely parameter configuration consistent with our
experimental setup gives a lifetime correction of -0.11 ns. We
perform a sensitivity analysis for the model parameters to obtain
an uncertainty in the lifetime of $\pm$ 0.1 \%.

%\bibliography{francium}

\begin{thebibliography}{10}

\bibitem{orozco02}
L.~A. Orozco,  in {\em Trapped Particles and Fundamental Physics,
Les Houches
  2000}, edited by S.~N. Atutov, R. Calabrese, and L. Moi (Kluwer Academic
  Publishers, Amsterdam, 2002).

\bibitem{simsarian98}
J.~E. Simsarian, L.~A. Orozco, G.~D. Sprouse, and W.~Z. Zhao,
Phys. Rev. A {\bf
  57},  2448  (1998).

\bibitem{grossman00b}
J.~M. Grossman, R.~P.~F. {III}, L.~A. Orozco, M.~R. Pearson, and
G.~D. Sprouse,
  Phys. Rev. A {\bf 62},  062502  (2000).

\bibitem{aubin03b}
S. Aubin, E. Gomez, L.~A. Orozco, and G.~D. Sprouse, Opt. Lett.
{\bf 28},  2055
   (2003).

\bibitem{aubin04}
S. Aubin, E. Gomez, L.~A. Orozco, and G.~D. Sprouse, Phys. Rev. A
{\bf 70},
  042502  (2004).

\bibitem{gomez04}
E. Gomez, S. Aubin, L.~A. Orozco, and G.~D. Sprouse, J. Opt. Soc.
Am. B {\bf
  21},  2058  (2004).

\bibitem{gomez05a}
E. Gomez, L.~A. Orozco, A.~P. Galvan, and G.~D. Sprouse, arXiv:
physics/0412073
  {\bf submitted to Phys. Rev.},  December  (2004).

\bibitem{wood99}
C.~S. Wood, S.~C. Bennett, J.~L. Roberts, D. Cho, and C.~E.
Wieman, Can. J.
  Phys. {\bf 77},  7  (1999).

\bibitem{guena03}
J. Gu\'{e}na, D. Chauvat, P. Jacquier, E. Jahier, M. Lintz, S.
Sanguinetti, A.
  Wasan, M.~A. Bouchiat, A.~V. Papoyan, and D. Sarkisyan, Phys. Rev. Lett. {\bf
  90},  143001  (2003).

\bibitem{johnson03}
W.~R. Johnson, M.~S. Safronova, and U.~I. Safronova, Phys. Rev. A
{\bf 67},
  062106  (2003).

\bibitem{ginges04}
J.~S.~M. Ginges and V.~V. Flambaum, Phys. Rep.  .

\bibitem{khriplovich97}
I.~B. Khriplovich and S.~K. Lamoreaux, {\em CP Violation Without
Strangeness:
  Electric Dipole Moments of Particles, Atoms and Molecules} (Springer-Verlag,
  New York, 1997).

\bibitem{oconnor84}
D.~V. O'Connor and D. Phillips, {\em Time Correlated Single Photon
Counting}
  (Academic, London, 1984).

\bibitem{hoeling96}
B. Hoeling, J.~R. Yeh, T. Takekoshi, and R.~J. Knize, Opt. Lett.
{\bf 21},  74
  (1996).

\bibitem{volz96b}
U. Volz and H. Schmoranzer, Phys. Scr. {\bf 65},  48  (1996).

\bibitem{gutterres02}
R.~F. Gutterres, C. Amiot, A. Fioretti, C. Gabbanini, M. Mazzoni,
and O.
  Dulieu, Phys. Rev. A {\bf 66},  024502  (2002).

\bibitem{heinzen97}
D.~J. Heinzen, private communication (unpublished).

\bibitem{aubin03a}
S. Aubin, E. Gomez, L.~A. Orozco, and G.~D. Sprouse, Rev. Sci.
Instrum. {\bf
  74},  4342  (2003).

\bibitem{zhao98}
W.~Z. Zhao, J.~E. Simsarian, L.~A. Orozco, and G.~D. Sprouse, Rev.
Sci.
  Instrum. {\bf 69},  3737  (1998).

\bibitem{diberardino98}
D. DiBerardino, C.~E. Tanner, and A. Sieradzan, Phys. Rev. A {\bf
57},  4204
  (1998).

\bibitem{bevington}
P.~R. Bevington and D.~K. Robinson, {\em Data Reduction and Error
Analysis for
  the Physical Sciences, Third edition} (McGraw-Hill, New York, 2002).

\bibitem{holstein47}
T. Holstein, Phys. Rev. {\bf 72},  1212  (1947).

\bibitem{beeler03}
M. Beeler, R. Stites, S. Kim, L. Feeney, and S. Bali, Phys. Rev. A
{\bf 68},
  013411  (2003).

\bibitem{knight80}
P. Knight, Opt. Commun. {\bf 32},  261  (1980).

\bibitem{marek80}
J. Marek and P. Munster, J. Phys. B {\bf 13},  1731  (1980).

\bibitem{safronova04}
M.~S. Safronova, C.~J. Williams, and C.~W. Clark, Phys. Rev. A
{\bf 69},
  022509  (2004).

\bibitem{johnson96}
W.~R. Johnson, Z.~W. Liu, and J. Sapirstein, At. Data Nucl. Data
Tables {\bf
  64},  279  (1996).

\bibitem{theodosiou84}
C.~E. Theodosiou, Phys. Rev. A {\bf 30},  2881  (1984).

\end{thebibliography}

\end{document}